\documentclass[]{aa}
\usepackage{graphicx}
\usepackage{txfonts}
\begin{document}
\title{Looking for Dust and Molecules in Nova V4743~Sagittarii 
\thanks{Based on observations collected with the Swedish ESO Submm Telescope at the European Southern Observatory, La Silla, Chile}}
\titlerunning{Dust and Molecules in V4743~Sgr}
\author{M. Nielbock 
        \and
        L. Schmidtobreick}
\offprints{M. Nielbock, \email{mnielboc@eso.org}}
\institute{European Southern Observatory, Casilla 19001, Santiago 19, Chile.}
\date{Received xxx xxx, xxx; accepted xxx xxx, xxx}
\abstract{We present 1.2~mm continuum images and spectral line observations
of CO(1--0) and SiO(3--2) rotational transitions of the recent nova V4743~Sgr. 
The nova is detected at 1.2~mm showing a variable millimetre emission.
Only upper limits of $T_{\rm A}^\star = 0.06$~K for CO and
$T_{\rm A}^\star = 0.03$~K for SiO could be derived. We discuss
the results in terms the nature of the millimetre emission favouring dust from
a phase before the recent outburst as the likely radiating source.
We also comment on the possibility of free--free emission
from the ionised shell as the source of the measured millimetre radiation.
}
\maketitle
\section{Introduction}
\object{V4743~Sgr} was discovered as a possible nova by
Katsumi Haseda (\cite{hase02}) at about 5~mag on September 20th, 2002. 
West (\cite{west02}) measured its position as 
$\rm R.A.(J2000) = 19^h01^m09\fs38$, 
$\rm Dec.(J2000) = -22^\circ00^\prime05\farcs9$
with an uncertainty of $0\farcs75$. Kato et al.
(\cite{kato+02}) confirmed that the object is an \ion{Fe}{ii}-class nova,
and measured the FWHM of the H$\alpha$ emission line of 2400 km\,s$^{-1}$. 
The light curves of AAVSO 2002 reveal it to be a very fast nova
undergoing a steep decline from its maximum and reaching three magnitudes 
below maximum within 15 days. This is usually understood as the transition
phase where dust formation might begin. 

The formation of dust has been suggested in many novae because of
a) the behaviour of their visual light curves which
indicate the increase of interstellar extinction at the beginning
of the transition phase, and b) the development of the infrared excess at 
the same time (see Bode \& Evans \cite{bode+89} for an overview).
The dust that forms during nova eruptions consists of a variety of 
mineralogies, and in some novae, simultaneous evidence is found for 
probably co-existing
carbon--rich and oxygen--rich condensates (Evans et al. \cite{evan+97}).

While most recent novae have been well observed in the optical and even 
infrared, hardly any observation exists at sub-millimetre or millimetre
wavelengths. Still, these measurements are valuable, because they allow to
investigate i) the expansion of the ionised shell from emission of free-free
radiation as a progenitor of the colder neutral matter, ii) the formation
of molecules like CO and other abundant
species of the interstellar medium (ISM), iii) the subsequent formation of
dust, which has a deep impact on our understanding of
the enrichment of the ISM and hence the formation of stars. 

The James Clerk Maxwell Telescope nova--monitoring group
detected V4743~Sgr with SCUBA (Ivison \cite{ivis02}). On October 3rd,
about 12 days after visual maximum,
they measured a flux density of $39\pm4$~mJy at 850~$\mu$m and determined
an upper limit of 225~mJy at 450~$\mu$m.

In this paper, we present observations of the 1.2~mm continuum emission
and the spectral line emission of CO(1--0) and SiO(3--2) of V4743~Sgr taken 
in a range between about 20 and 60 days after the visual maximum. 
We compare the 1.2~mm emission with Ivison's 450 and 850~$\mu$m 
data and discuss the implications for the origin of the millimetre emission.

\section{Observations and data reduction}
The observations were carried out at the SEST on La Silla, Chile.
Millimetre continuum observations at 1.2~mm 
were taken using the SEST Imaging Bolometer Array (SIMBA) 
operating at 250~GHz with a FWHM bandwidth of 90~GHz. The maps were 
produced with the fast-scanning mode (Reichertz et al. \cite{reic+01})
without a wobbling secondary mirror. The correlated sky noise was
eliminated during the data reduction process with 
MOPSI\footnote{MOPSI has been developed by Dr.~R.~Zylka, 
IRAM, Grenoble, France}. 
The zenith opacity at 1.2~mm was determined with
skydips, and ranged between $\tau_0 = 0.18$ and $0.44$. 
Flux calibration was achieved with a series of Uranus maps. 
We estimate it to be accurate to 10\%.

\begin{figure*}
\rotatebox{-90}{\resizebox{!}{18cm}{\includegraphics{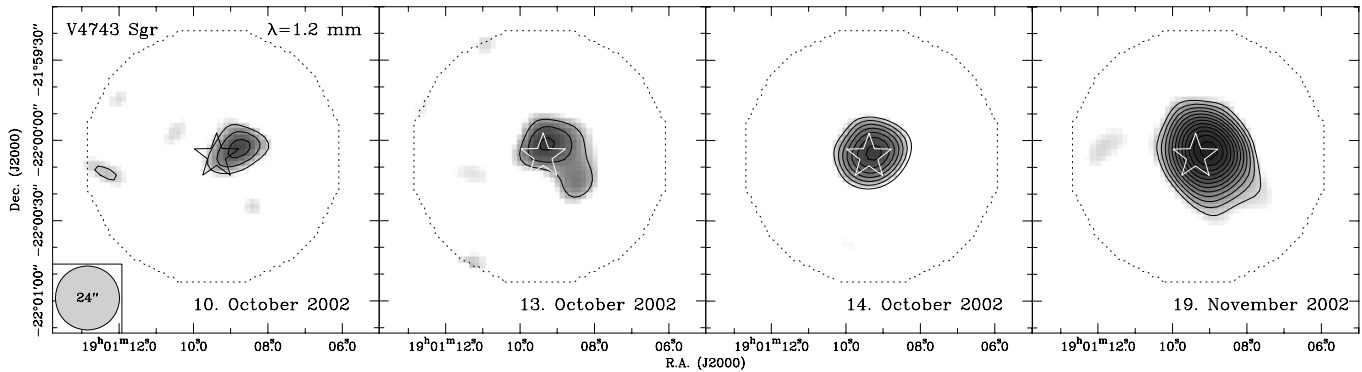}}}
\caption{\label{maps} The individual maps of V4743~Sgr at 1.2~mm
continuum emission. The position of the nova as determined from optical
observations is marked as a star. The beam size (lower left corner)
is too large as that the structure of the source can be regarded as real.}
\end{figure*}

Spectra of the rotational transitions of CO(1--0) at 115.3~GHz 
and SiO(3--2) at 127.1~GHz were acquired using the 
SESIS~100/150 heterodyne receiver providing a FWHM beam size of about
60\arcsec. We used the acousto-optical spectrometer (AOS) at low resolution
(LRS) in order to achieve the largest possible velocity coverage. 
Two spectra for each molecule were produced, one at the centre velocity
of 0~km s$^{-1}$, and another with an offset of +800~km s$^{-1}$ (LSR).
The separation of each two of the 1440 spectrometer channels is 0.7~MHz being 
equivalent to 1.8~km s$^{-1}$ and 1.6~km s$^{-1}$, respectively. Dual beam
switching was chosen for the rejection of the sky emission and system
imbalances. A summary of the observational parameters is given in
Tables~\ref{obs_simba} and \ref{obs_specs}.

\begin{table}
\caption{\label{obs_simba}
Details of the continuum observations with SIMBA at \mbox{$\lambda =
1.2$~mm}:
date, number of exposures, individual integration time, and atmospheric
zenith opacity.}
\begin{tabular}{ l l l l }
\hline
\hline
\multicolumn{1}{c}{Date (2002)} & 
\multicolumn{1}{c}{No$_{\rm exp}$}& 
\multicolumn{1}{c}{$t_{\rm exp}$ [s]} &
\multicolumn{1}{c}{$\tau_0$}\\
\hline
10th Oct & 14 & 264 & 0.18 -- 0.22 \\
13th Oct & 10 & 264 & 0.22 -- 0.37 \\
14th Oct & 17 & 264 & 0.43 -- 0.44 \\
19th Nov & 12 & 264 & 0.18 -- 0.22 \\
\hline
\end{tabular}
\end{table}

\begin{table}
\caption{\label{obs_specs}
Details of the spectral line observations: date, frequency, 
molecular transition, central velocity (LSR), system temperature,
and integration time.}
\tabcolsep3.5pt
\begin{tabular}{ l l l l l l }
\hline
\hline
\multicolumn{1}{c}{Date (2002)} &
\multicolumn{1}{c}{$\nu$ [GHz]} &
\multicolumn{1}{c}{Molecule} &
\multicolumn{1}{c}{$v_{\rm LSR}$ [km s$^{-1}$]} &
\multicolumn{1}{c}{$T_{\rm sys}$ [K]} &
\multicolumn{1}{c}{$t_{\rm obs}$ [s]} \\
\hline
15th Oct & 115.3 & CO(1--0)  &    0 & 446 & 600 \\
15th Oct & 115.3 & CO(1--0)  & +800 & 531 & 450 \\
15th Oct & 127.1 & SiO(3--2) &    0 & 231 & 600 \\
15th Oct & 127.1 & SiO(3--2) & +800 & 271 & 450 \\
\hline
\end{tabular}
\end{table}

\section{Results}
V4743~Sgr was detected in all 1.2~mm continuum measurements
with a S/N of at least 4. A confusion with another source
is unlikely. Apart from rare phenomena like novae and supernovae,
only quasars and similar background sources show a strong
variation as witnessed in our study. 30 per $\sq^\circ$ are expected
for the detection limit of our observations (Blain et al.~\cite{blain98}).
This gives about 0.02 sources for our map coverage.

The individual 1.2~mm maps are presented in Fig.~\ref{maps}. Note that the
structures inside the source are small in comparison to the beam size (lower
left corner) and are therefore probably not real. The flux densities have been 
determined in two ways:
a) by integrating over a 1\arcmin\ aperture while subtracting the background 
($S_{\rm ap}$), and b) by fitting a two--dimensional Gaussian to the source 
and integrating the flux within ($S_{\rm G}$). Table~\ref{flux} lists both
values for all epochs as well as the 1$\sigma$ rms noise as the photometric
uncertainty. A millimetre light curve is plotted in Fig.~\ref{lc}. We used
$S_{\rm G}$ for a further analysis. The point-like source should have a Gaussian
shape, but atmospheric fluctuations distort the images of weak sources.
Therefore, a simple aperture photometry might be misleading. In addition, the
integration over a Gaussian is more comparable to on-off measurements.

The spectral line observations of the CO(1--0) and the SiO(3--2) rotational
transitions only provided upper limits of $T_{\rm A}^\star = 0.06$~K for the
CO and $T_{\rm A}^\star = 0.03$~K for the SiO measurements within the covered
velocity range.

\begin{table}
\caption{\label{flux}
Flux densities at 1.2~mm as determined via aperture photometry 
($S_{\rm ap}$) within a diameter of 1\arcmin\  and via integration
over a fitted 2-dim Gaussian ($S_{\rm G}$) with their uncertainty
are listed for each date.}
\begin{tabular}{ l r r }
\hline
\hline
\multicolumn{1}{c}{HJD [2452000 +]} &
\multicolumn{1}{c}{$S_{\rm ap}$ [mJy]} &
\multicolumn{1}{c}{$S_{\rm G}$ [mJy]} \\
\hline
558.406 &  $16 \pm 4$ &   $21 \pm 4$ \\
561.271 &  $65 \pm 9$ &   $52 \pm 9$ \\
562.342 &  $30 \pm 5$ &   $39 \pm 5$ \\
598.483 & $130 \pm 8$ &  $142 \pm 5$ \\
\hline
\end{tabular}
\end{table}

\section{Discussion}
Millimetre continuum radiation from novae can arise from thermal dust emission
(e.g. Evans et al.~\cite{evan+97}) or as free--free emission from hot 
ionised gas (e.g. Ivison et al.~\cite{ivis+93}). The decision
on which is the dominant emission source is made on the shape of the
spectral energy distribution (SED), i.e. the slope $\alpha$ of its 
Rayleigh--Jeans tail. The SED of dust emission is usually
described by a modified blackbody law. The flux density $S_\nu$ at a
frequency $\nu$ is given by
\begin{equation}
S_\nu = \Omega\, B_\nu(T) \left(1-\exp\left(-\tau_\nu\right)\right)\\
\tau_\nu = \left(\nu/\nu_{\rm c}\right)^\beta\\
\label{e:greybody}
\end{equation}
where $\Omega$ is the solid angle of the emitting source,
$T$ the dust temperature, $\tau_\nu$ the optical depth, and $B_\nu(T)$ the
Planck function. The optical depth is a function of the frequency with an
exponent $\beta = \alpha -2$,  which often is referred to as the submm 
emissivity index,
and $\nu_{\rm c}$ is the critical frequency, i. e. the frequency where
$\tau_\nu = 1$. Values between $\beta = 1$ and $\beta = 2.5$ are found
for different dust grains. Free--free emission instead yields a flat
SED ($\alpha = -0.1$) for optically thin and $\alpha = 2$ for optically thick
media (e.g. Chen et al.~\cite{chen95}). 

\begin{figure}
\centering
\resizebox{\hsize}{!}{\includegraphics{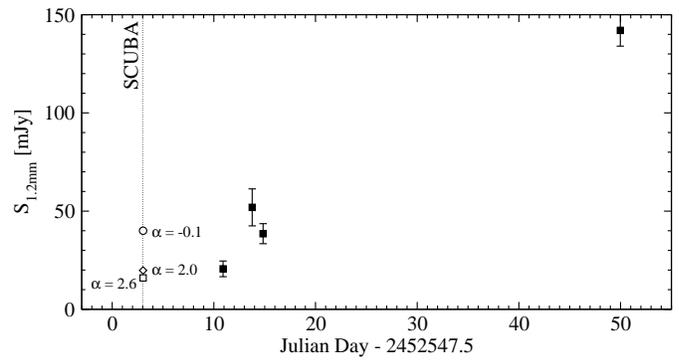}}
\caption{\label{lc} Millimetre light curve of V4743~Sgr. The four
rightmost squares mark the observed 1.2~mm flux densities 
$S_{\rm G}$. The intensities at the date of the SCUBA
measurement (indicated as a vertical dotted line) have been extrapolated from
the SCUBA data at 850~$\mu$m for different submm slopes $\alpha$.}
\end{figure}

Unfortunately, no simultaneous measurements at mm and submm exist for V4743~Sgr.
Therefore, we calculated 1.2~mm values for the time of the SCUBA 
measurements for the cases of free--free and dust emission,
and included them in Fig.~\ref{lc}. The steepest submm slope that is
consistent with the SCUBA data is $\alpha = 2.6$. If the mm emission
were due to free--free radiation from an expanding ionised shell, it would be
difficult to explain the decreasing flux density in the beginning.
Hence, free--free emission seems to be an unlikely source,
and we favour the dust interpretation. 
Additional evidence comes from the optical light-curves provided by 
AAVSO. Coincidently with the increase in the submm, they show a
rise of the $R$-band intensity, while values at shorter wavelengths
remain constant.

The formation of dust in novae is usually observed by an increasing IR emission 
accompanied by a prominent dip in the visual light curve of the nova which
can be explained by extinction introduced by newly formed dust. Such an
extinction dip has not been observed for V4743~Sgr. Together with the lack of
any emission from CO and SiO molecules, we conclude that the millimetre
emission does not originate from newly formed but old dust. It might be
either a remnant from the pre-CV phase or a former nova outburst.
In that case, V4743~Sgr would actually be a recurrent nova. 
A similar interpretation
was made for V4444 Sgr which showed a rise in the IR without any dip in
the optical light curve (Venturini et al. \cite{vent+02}). If we follow the
argument of Kawabata et al.~(\cite{kawa00}), we can use Eq.~(6) of Clayton \&
Wickramasinghe~(\cite{clay76}) to verify that dust cannot be present inside
a radius of $\sim$50~AU from the nova. The largest possible radius of the
expanding nova shell of V4743~Sgr can be derived by extrapolating the
expansion speed measured from the H$\alpha$ line width of 2400~km s$^{-1}$
for 23 days to the first millimetre maximum. This yields 4.8 10$^9$~km or
32~AU. Hence, the dust cannot originate from the recent outburst. The
presence of pre-existing dust around novae has been shown for several 
more cases (e.g.  Malakpur~\cite{mala77}).
\begin{equation}
M_{\rm dust}=\frac{S_\nu\, D^2}{\kappa_\nu\, B_\nu(T)}
\label{e:mass}
\end{equation}

Eq.~(\ref{e:mass}) allows to estimate the  mass of the surrounding
dust, if $D$ the distance to the source, $\kappa_\nu$ the mass absorption 
coefficient at the measured frequency $\nu$, and $T$ the dust temperature are
known. $\kappa_\nu$ can differ at least by a factor of 10, depending on the
local environment. In our case, we adopt a value of
$\kappa_{250}=2$~cm$^2$ g$^{-1}$, which is typical for circumstellar dust
(Kr\"ugel \& Siebenmorgen \cite{kruegel94}) consisting of relatively small
grains (Kawabata et al.~\cite{kawa00}).

The determination of the distance $D$ of CVs is generally quite difficult,
although some methods exist to estimate the distance modulus of novae.
Firstly, the absolute visual magnitude of a nova 15 days after the maximum
is supposed to be similar for all novae, i.e. $M_V(15) = -5.69 \pm 0.14$~mag
(Capaccioli et al.~\cite{capa+90}, and others herein). Secondly, the absolute 
visual magnitude of a nova at its maximum is related to the time $t_2$, in
which the nova descends by 2~mag (Cohen~\cite{cohe88}).
We estimated the distance modulus of V4743~Sgr with both methods and 
get $(m-M) = 13.9 \pm 0.4$ which yields a maximum possible distance of 6~kpc. 
However, the crucial point is to measure the interstellar extinction towards
the nova. The region looks very patchy and \object{HD~176537}, the closest
star with a measured extinction, still lies 40\arcmin~away from the nova
itself and hence its value $A_V = 1.42$~mag (Guarinos~\cite{guar92}) might
differ quite a lot from that of V4743~Sgr. Additionally, its heliocentric
distance is only about 100~pc. Assuming that HD~176537 and the nova are
sharing the same interstellar environment, and scaling the extinction value
accordingly, we estimate the distance of V4743~Sgr to be around $D = 400$~pc
with $A_V = 5.7$ mag.
Using the $J-H$ and $H-K$ colours of the 2Mass point sources in a vicinity of
1\arcmin~ in comparison with intrinsic colours from Bessel \& Brett
(\cite{bess+88}), we estimate the average extinction as $A_V = 3.4\pm 0.5$~mag
which yields $D = 1200$~pc for the nova. We want to point out that these
values are very speculative and have to be regarded with caution. For this
reason, we relate the mass estimates to a standard distance of 100~pc.

A well established estimate
of the dust temperature is crucial for deriving a proper dust mass from
millimetre observations. The present data are insufficient to achieve this,
and temperature determinations are sparse throughout the literature. Evans
et al.~(\cite{evan+97}) quote a dust temperature of some 750~K for
\object{V705~Cas} after 250--300 days after the nova eruption, and Bode \&
Evans~(\cite{bode+89}) give values between 700 and 800~K for
\object{NQ~Vul} between 80 and 240 days after the outburst. However,
these values have been derived from infrared measurements during the ionisation
phase and therefore
do not trace the coolest dust involved. Consequently, they may not be
appropriate for our purposes.

On the other hand, Lynch et al.~(\cite{lynch+01}) derive a
temperature of 280~K for the cool dust component in \object{V445~Pup}.
Beck et al.~(\cite{beck+90}) present models of nova environments,
in which one of the most important ingredients for dust formation, \ion{C}{i},
attains a temperature of about 100~K already soon after the nova eruption. 
We therefore use typical dust temperatures of 100 and 300~K for
the mass determination. The so derived dust masses are listed
in Table~\ref{t:mass} for 100 and 300~K, respectively. As discussed above,
they are scaled to a distance of 100~pc.  

We determined column densities of the surrounding dust from the derived
masses. The projected area in the line of sight was calculated from the lowest
possible dust radius, i.e. 50~AU (see above). Since the area can be
larger than this, these column densities are only an upper limit and are also
listed in Table~\ref{t:mass}.

\begin{table}
\caption{\label{t:mass}
Dust masses and column densities as calculated from
Eq.~(\ref{e:mass}). We assume a standard distance of 100~pc for estimating
the mass.}
\tabcolsep5pt
\begin{tabular}{ c c c c c }
\hline
\hline
HJD  &
\multicolumn{2}{c}{$M_{\rm dust}\left(\frac{D}{100~{\rm pc}}\right)^{-2}$ [$M_\odot$]} &
\multicolumn{2}{c}{$N\left(\frac{D}{100~{\rm pc}}\right)^{-2}$ [cm$^{-2}$]} \\
$[2452000 +]$ &$T=100$~K   &  $T=300$~K &
$T=100$~K   &  $T=300$~K \\
\hline
558.406 & 2.8 10$^{-6}$ & 8.9 10$^{-7}$ & 7.0 10$^{20}$ & 2.2 10$^{20}$ \\
561.271 & 6.9 10$^{-6}$ & 2.2 10$^{-6}$ & 1.7 10$^{21}$ & 5.5 10$^{20}$ \\
562.342 & 5.2 10$^{-6}$ & 1.7 10$^{-6}$ & 1.3 10$^{21}$ & 4.2 10$^{20}$ \\
\hline
\end{tabular}
\end{table}

If we assume the distance estimates from above, we get average dust masses of about
$5\,10^{-5}~M_\odot$ for $D=400$~pc and $5\,10^{-4}~M_\odot$ for $D=1200$~pc.
The referring column densities are $1\,10^{22}$~cm$^{-2}$ and 
$1\,10^{23}$~cm$^{-2}$, respectively. We recall that a distance
of 400~pc is probably a gross underestimate. We therefore
think that the actual values are more likely to be found at the higher end. 

If the flux variation were due to forming dust, the dust formation rate
$\dot{M}_{\rm dust}$ could be estimated from the dust masses
between the first two measurements (2.9 days). For the distance estimates
from above we find $3.5\,10^{23}$ and $3.1\,10^{24}$~g s$^{-1}$, respectively.
We compare these dust formation rates with the results from two other novae,
\object{Nova Serpentis 1978} (\object{LW~Ser}) and \object{Nova Cassiopeiae
1993} (\object{V705~Cas}). Gehrz et al.~(\cite{gehrz+80}) measured the mass of
the dust produced during the outburst of LW~Ser from infrared observations.
The infrared data show a steep rise to a maximum within 41 days, which we take
as the time of the dust formation. The average dust formation rate is therefore
$1.8\,10^{20}$~g s$^{-1}$. Evans et al.~(\cite{evan+96}, \cite{evan+97})
show a visual light curve of V705~Cas with a prominent extinction dip of 29
days, which we define as the dust formation phase. The average dust formation
rate in this case yields $7.9\,10^{18}$~g s$^{-1}$. Our rates, if real, would
be at least $2000$ times higher. A factor of about $17000$, however, seems more
likely. We interpret this discrepancy again as an indication that the radiating
dust was most likely not produced after the nova eruption.

\begin{equation}
\tau_\nu = N\,m_{\rm d}\,\kappa_\nu
\label{e:tau}
\end{equation}

From Eq.~(\ref{e:tau}) we determine an average optical depth of the
medium at 1.2~mm of $<\!\!\tau_{250}\!\!>\ = 9\,10^{-3}$, where $N$ is the 
column density and $m_{\rm d}=4.5\,10^{-26}$~g is the mass of a dust particle. 
This is taken as 1\% of the particle mass of the ISM, where we correct
for the typical mass abundance of helium.
With Eq.~(\ref{e:greybody}) we can calculate the frequency for
which the medium becomes optically thick. Transforming this frequency into
a wavelength we get $\sim$175~$\mu$m for
$\beta = 0.6$. The size of the bulk of the involved scattering dust particles
should be of the order of $a=\lambda/2\pi$, i.e. $38~\mu$m.

\section{Summary}
V4743~Sgr was detected at 1.2~mm, the flux density varying strongly in
time. A local intensity maximum has been found around day 23 after the 
visual maximum. It continued to rise steeply afterwards. The spectral line
observations yielded only upper limits for CO(1--0) and SiO(3--2).
The dominant emission source is found to be heated dust rather than
free--free emission.
We have roughly estimated the distance of the nova as $1200\pm 300$~pc.
Using this value, the dust mass at the first maximum ranges between
$3\,10^{-4}~M_\odot$ and $1\,10^{-3}~M_\odot$ depending on its temperature.
The high dust mass, the accordingly high formation rate, the absence of
molecules, the lack of any dip in the visual light curve while the $R$-band
magnitude rises simultaneously with the mm flux, and the small
shell radius favour the idea that the
observed dust has already been present in the system as a remnant of either
the pre-CV common envelope phase or a former nova outburst.
Dust grains were estimated to have sizes around 38~$\mu$m.

\begin{acknowledgements}
In this research, we have used, and acknowledge with 
thanks, data from the AAVSO International Database, 
based on observations submitted to the AAVSO by variable 
star observers worldwide.
This publication makes furthermore use of data products from the 
Two Micron All Sky
Survey, which is a joint project of the University of Massachusetts and the
Infrared Processing and Analysis Center/California Institute of Technology,
funded by the National Aeronautics and Space Administration and the National
Science Foundation. We also thank Andreas Lundgren and Dr.~Lars-\AA ke
Nyman for carrying out the November observation.
\end{acknowledgements}

\end{document}